\documentclass[preprint,showpacs,preprintnumbers,amsmath,amssymb]{revtex4}

\usepackage{graphicx}\graphicspath{./graph/}
\usepackage{bm}
\usepackage{color}
\usepackage{verbatim}
\usepackage{amssymb}
\usepackage{amsfonts}
\usepackage{latexsym}
\usepackage{epstopdf}

\definecolor{red}{rgb}{1,0,0}
\definecolor{green}{rgb}{0,1,0}
\definecolor{blue}{rgb}{0,0,1}

\newcommand{\be}{\begin{equation}}
\newcommand{\ee}{\end{equation}}
\newcommand{\bea}{\begin{eqnarray}}
\newcommand{\eea}{\end{eqnarray}}
\newcommand{\bdm}{\begin{displaymath}}
\newcommand{\edm}{\end{displaymath}}

\begin{document}

\title{Arrow, Hausdorff, and Ambiguities in the Choice of
Preferred States in Complex Systems}
\author{T. Erber$\dagger$ and M. J. Frank$\ddagger$}
\affiliation{Department of Physics Illinois Institute
of Technology, Chicago, USA$\dagger$\\
Department of Applied Mathematics Illinois Institute
of Technology, Chicago, USA$\dagger \, \ddagger$\\
Department of Physics University of Chicago,
Chicago, USA$\dagger$}

\begin{abstract}
Arrow's `impossibility' theorem asserts that there are no satisfactory methods of aggregating individual preferences
into collective preferences in many complex situations. This result has ramifications in economics, politics, i.e.,
the theory of voting, and the structure of tournaments. By identifying the objects of choice with mathematical sets,
and preferences with Hausdorff measures of the distances between sets, it is possible to extend Arrow's arguments from a
sociological to a mathematical setting. One consequence is that notions of reversibility can be expressed in terms of the
relative configurations of patterns of sets.
\end{abstract}


\maketitle

\section{1. Arrow's Impossibility Theorem}
\label{arrow}
One of the most significant 'no-go' results discovered in the twentieth century is Arrow's theorem concerning the
impossibility of devising satisfactory methods for aggregating individual preferences into collective preferences in many
complex situations \cite{arrow}. Arrow's arguments have given rise to a voluminous literature with wide ranging applications
in economics, politics, and the organization of tournaments. The feasibility of further extending these results to areas of
mathematics and physics depends on finding appropriate counterparts to Arrow's objects of choice, i.e., various social states,
and reinterpreting the concept of preference in quantitative terms. One possible approach is to identify the objects of choice
with mathematical sets, and to relate the associated preferences to Hausdorff's asymmetric distances between sets.

To be specific, let $A, B, C,$ etc. denote the objects of choice, and represent the relations of preference by means of the symbols $\succ$
and $\prec$ : consequently a formula such as $A \succ B$ signifies that $A$ is preferred over $B$, and similarly the expressions $B \succ A$
and $A \prec B$ both mean that $B$ is preferred to $A$. Suppose now that there are three individuals labeled 1, 2, 3, each of which can
choose between the two alternatives $A$ and $B$. Then, irrespective of which combination of preferences  -~-~-~ out of a total of eight
possibilities -~-~-~ occurs, there is a reasonable way of aggregating the individual preferences into a collective preference. For instance,
the pattern
\be
1:~  A \succ B\ ;\quad 2:~ A \succ B\ ;\quad 3:~ B \succ A     \label{choices3}
\ee
indicates the overall preference for the choice of $A$ in virtue of majority rule.

Arrow's arguments become relevant in the slightly more complicated situation where there are three individuals who can choose among three alternatives $A,~B,$ and $C$. Although in this case there are many combinations of preferences that can be aggregated into collective preferences with the help of plausible schemes such as majority rule, there are some recalcitrant outliers whose antecedents date back more than two centuries \cite{condorcet}. Suppose, for instance that the first individual orders the objects of choice in the sequence
\be
1:\quad A \succ B \succ C \succ A\ .     \label{sec1}
\ee
This is an instance where the preferences of $A$ over $B$ and $B$ over $C$ do not imply that $A$ is preferred over $C$. Sequences of this kind are said to be intransitive \cite{tarski}. The ambiguity is heightened still further if the remaining two individuals order their preferences in a cyclic version of Eq.(\ref{sec1}), viz.
\bea
&&2:\quad   B \succ C \succ A \succ B\ , \label{sec2}\\
&&3:\quad   C \succ A \succ B \succ C\ , \label{sec3}
\eea
Clearly majority rule can't achieve a consensus in this situation. Experience shows that even more elaborate voting schemes fail to extract a fair and favored
choice from such sets of intransitive preferences. Arrow's point is that this is not a matter of ingenuity but rather an impossibility:  in some complex situations
there is simply no reasonable method for aggregating individual preferences into a collective preference. The reasoning is based on making precise the notions of `fair' and
`reasonable'.

Arrow lists four criteria that ought to be satisfied by any acceptable voting procedure:
\begin{description}
\item{1.} Local and global harmony. Suppose that in a bloc of voters every individual
 has the same preference, say, $A$ over $B$ ($A \succ B$), then the collective preference of the entire group is also $A$ over $B$.
 \item{2.} All choices are possible. Every individual can in principle choose among all available alternatives and these can be
 ordered in every possible sequence of preferences.
 \item{3.} Independence of irrelevant alternatives. The set of choices, $A,~B,$ etc., available to every individual constitutes an
 environment of admissible options. All methods of aggregating individual preferences into a collective preference then must be
 independent of any choice that lies outside of the environment.
 \item{4.} Non-dictatorship. The collective preferences of a group of individuals are not to be determined solely by the preferences
 of a single individual.
  \end{description}

 At first sight these innocuous propositions appear to be part of any reasonable voting scheme, but more than sixty years ago Arrow
 discovered an astonishing twist -~-~-~ these propositions are actually incompatible! This basic flaw is the reason that it is impossible
 to devise a general method for aggregating individual preferences into a collective preference applicable under all circumstances.
 The technical details of the proof are given in \cite{arrow}. A more mathematically oriented treatment is presented in \cite{kelly}.

 \section{2. The Hausdorff distance between sets}
Let $S$ be a planar set composed of the points $s_i,\ 1\le i\le N_S$, and $T$ be another set in the same plane composed
of the points $t_j,\ 1\le j\le N_T$. Further, let $d(s_i,t_j)$ denote the ordinary Euclidean distance between $s_i$ and $t_j$. Then the Hausdorff distance from the set $S$ to the set $T$ is given by the radius of the smallest disk
centered at any point of $S$ that also includes at least one point of $T$ \cite{hausdorff}. This definition corresponds to the expression
\be
\delta_H(S \rightarrow T)=\sup_{s_i \in S}\ \inf_{t_j \in T} d(s_i,t_j)\ .          \label{delstot}
\ee
In a similar fashion the Hausdorff distance from the set $T$ to the set $S$ is given by the radius of the smallest disk
centered at any point of $T$ that also includes at least one point of $S$. This definition corresponds to a formula
analogous to (\ref{delstot}),
\be
\delta_H(T \rightarrow S)=\sup_{t_j \in T}\ \inf_{s_i \in S} d(s_i,t_j)\ .          \label{delttos}
\ee
In general, these distances depend both on the configurations as well as the relative positions of the sets $S$ and $T$. Consequently, the directed Hausdorff distances $\delta_H(S \rightarrow T)$ and $\delta_H(T \rightarrow S)$ may be
unequal, even though the underlying Euclidian metric $ d(s_i,t_j)$ is symmetric. This is the essential property that
furnishes a link between the concepts of choice and preference and the mathematical notion of a distance between sets

\begin{figure}
\centering
 \includegraphics[width=\linewidth]{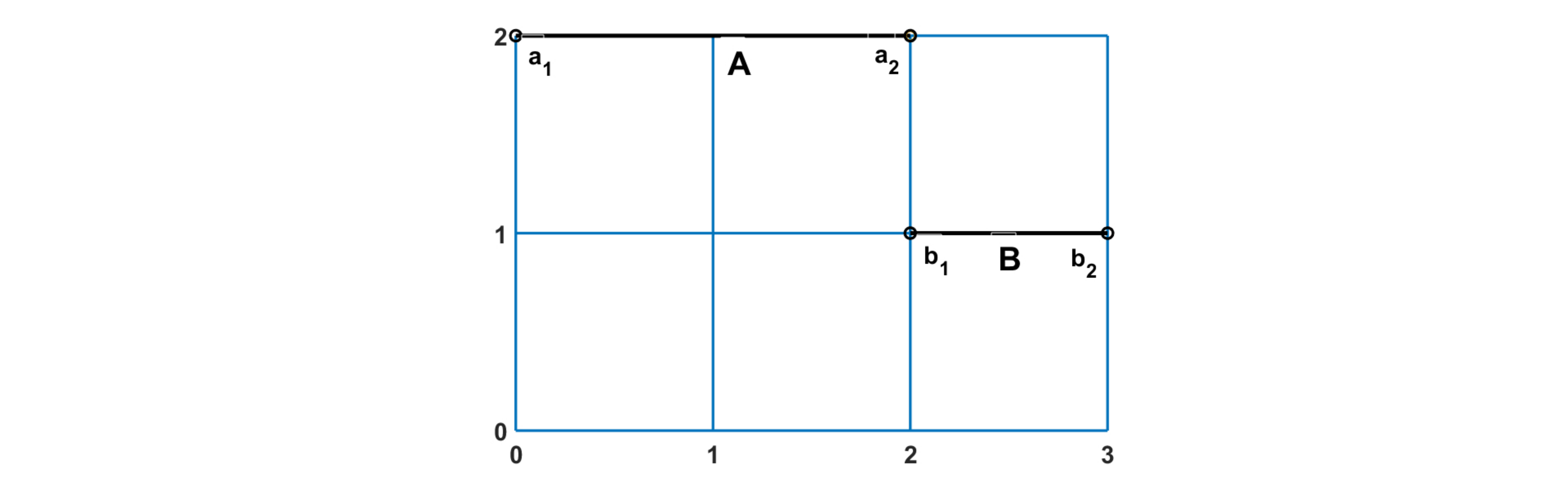}
\caption{Arrangement of the point sets $A$ and $B$ over a grid of unit squares}
\label{fig:1}       
\end{figure}

The transition from Arrow to Hausdorff can be illustrated with the aid of several very simple examples. Suppose that $A$ and $B$ are both two-point sets with the elements $a_1,~a_2 \in A$, and $b_1,~b_2 \in B$, arranged over a grid of unit squares as shown in Figure \ref{fig:1}. In this case the basic distance definition (\ref{delstot}) reduces to the simpler form
\be
\delta_H(A \rightarrow B) = \max\big\{\min_i ~d(a_1,b_i),\ \min_i~ d(a_2,b_i)\big\}\ .  \label{del2ptab}
\ee
The four Euclidean distances between the points in the two sets can then be displayed in the form of a $2\times2$ array
\be
\bordermatrix{& b_1 & b_2 \cr
a_1 & \sqrt{5} & \sqrt{10} \cr
a_2 & 1 & \sqrt{2} \cr}
 \label{array1}
\ee
Inserting these numbers into (\ref{del2ptab}) one obtains
\be
\delta_H(A \rightarrow B) = \max\big\{ \min(\sqrt{5},\sqrt{10}),~\min(1,\sqrt{2}) \big\}\ ,   \label{delab1}
\ee
or
\be
\delta_H(A \rightarrow B) = \max\big\{ \sqrt{5},\ 1\big\}= \sqrt{5}\ .   \label{delab2}
\ee
The reverse Hausdorff distance $\delta_H(B \rightarrow A)$ can then be computed from the analog of (\ref{del2ptab}), viz.
\be
\delta_H(B \rightarrow A) = \max\big\{\min_i ~d(b_1,a_i),\ \min_i~ d(b_2,a_i)\big\}\ .  \label{del2ptba}
\ee
Again, inserting the numbers, the result is
\be
\delta_H(B \rightarrow A) = \max\big\{ \min(\sqrt{5},1),~\min(\sqrt{10},\sqrt{2}) \big\}\ ,   \label{delba1}
\ee
or
\be
\delta_H(B \rightarrow A) = \max\big\{ 1,\ \sqrt{2}\big\}= \sqrt{2}\ .   \label{delba2}
\ee
Clearly, the numerical differences between the two Hausdorff distances are due to the fact that (\ref{delab2})
corresponds to a `max - min by rows' algorithm whereas (\ref{delba2}) corresponds a `max - min by columns' algorithm.
The two parallel interpretations of the expression $A \succ B$ are now complete: in Arrow's language this means that the social state $A$ is preferred over the social state $B$; in Hausdorff's terminology $A$ and $B$ are sets
whose configurations and relative positions imply that the Hausdorff distance from $A$ to $B$ is greater than the distance from $B$ to $A$.

\begin{figure} [htb]
\centering
 \includegraphics[width=\linewidth]{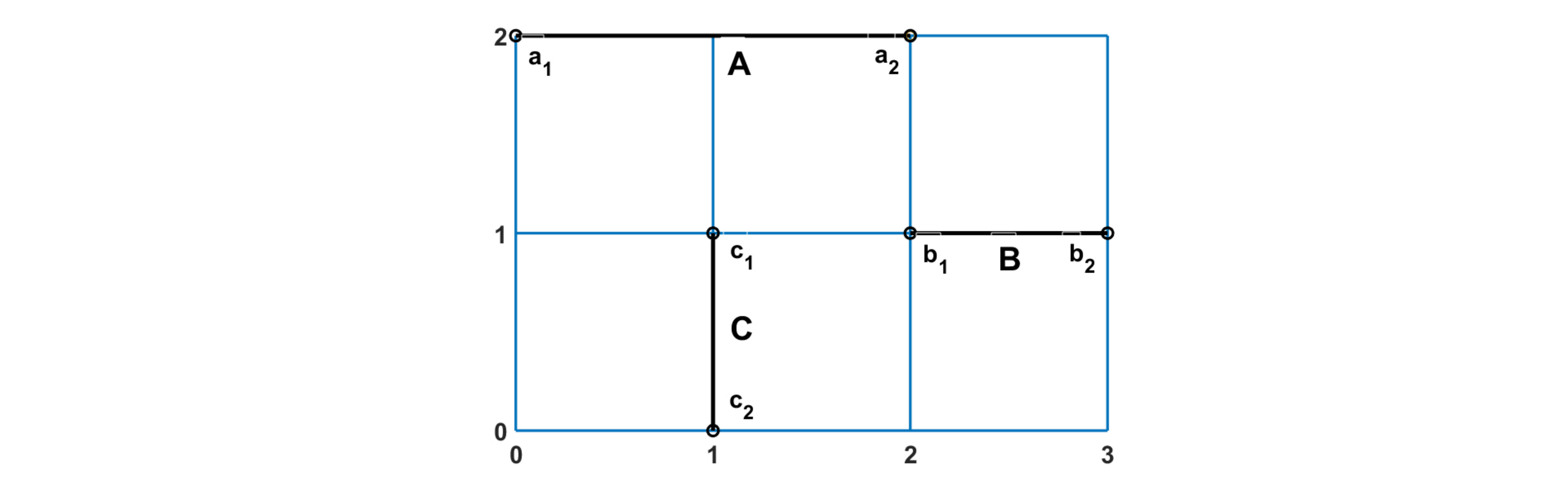}
\caption{Arrangement of three point sets $A,~B,~C$ whose mutual Hausdorff distances satisfy the inequalities
in Eq.(\ref{uni}).}
\label{fig:2}       
\end{figure}

The next increment of complexity is the Condorcet triplet Eq.(\ref{sec1}). This intransitive sequence of preferences can also be associated with a pattern of sets.  Figure \ref{fig:2} shows one possible arrangement of  three seta $A,~B,~C$ whose relative Hausdorff distances mirror the preference rankings in Eq.(\ref{sec1}). It is easy to confirm this numerically since the $A$ and $B$ sets are exactly the same as in Figure \ref{fig:1}; and it merely remains to evaluate the new Hausdorff  distances $\delta_H(B \rightarrow C)$ and $\delta_H(C \rightarrow A)$. Following the steps of the prior calculations in Eqs.(\ref{del2ptab})-(\ref{array1}), we first list the Euclidian distances between the points $B$ and $C$ in an array
\be
\bordermatrix{& c_1 & c_2 \cr
b_1 & 1 & \sqrt{2} \cr
b_2 & 2 & \sqrt{5} \cr}
\label{array2}
\ee
The corresponding Hausdorff distance is then
\be
\delta_H(B \rightarrow C) = \max\big\{ \min(1,\sqrt{2}),~\min(2,\sqrt{5}) \big\}\ ,   \label{delbc1}
\ee
or
\be
\delta_H(B \rightarrow C) = \max\big\{ 1,\ 2\big\}= 2\ .        \label{delbc2}
\ee
The reverse Hausdorff distance is given by
\be
\delta_H(C\rightarrow B)=\max\big\{\min(1,2),\ \min(\sqrt{2},\sqrt{5})\big\}\ ,  \label{delcb1}
\ee
or
\be
 \delta_H(C \rightarrow B) = \max\big\{ 1,\ \sqrt{2}\big\}= \sqrt{2}\ .    \label{delcb2}
\ee
Similarly the distance between $C$ and $A$ can be inferred from the array
\be
\bordermatrix{& c_1 & c_2 \cr
a_1 & \sqrt{2} & \sqrt{5} \cr
a_2 & \sqrt{2} & \sqrt{5} \cr}
\label{array3}
\ee
Specifically,
\be
\delta_H(A \rightarrow C) = \max\big\{ \min(\sqrt{2},\sqrt{5}),~\min(\sqrt{2},\sqrt{5}) \big\}\ ,   \label{delac1}
\ee
or
\be
\delta_H(A \rightarrow C) = \max\big\{ \sqrt{2},\ \sqrt{2}\big\}= \sqrt{2}\ .    \label{delac2}
\ee
The last distance is given by
\be
 \delta_H(C \rightarrow A) = \max\big\{ \min(\sqrt{2},\sqrt{2}),~\min(\sqrt{5},\sqrt{5}) \big\}\ ,   \label{delca1}
\ee
or
\be
\delta_H(C \rightarrow A) = \max\big\{ \sqrt{2},\ \sqrt{5}\big\}= \sqrt{5}\ .    \label{delca2}
\ee
The equivalences between the preference rankings in Eq.(\ref{sec1}) and the distance inequalities implicit in Figure 2 can
now be listed in a unified form:
\bea
&& A \succ B ~\Leftrightarrow~ \delta_H(A \rightarrow B)=\sqrt{5}\ > \ \sqrt{2} = \delta_H(B \rightarrow A)\nonumber\\
&& B \succ C ~\Leftrightarrow~ \delta_H(B \rightarrow C)=\ 2\ \ > \ \sqrt{2} = \delta_H(C \rightarrow B)\label{uni}\\
&& C \succ A ~\Leftrightarrow~ \delta_H(C \rightarrow A)=\sqrt{5}\ > \ \sqrt{2} = \delta_H(A \rightarrow C)\nonumber
\eea

\section{3. Reversions}
In ordinary particle dynamics a reversal of motion is usually effected by a reversal of all velocity components. This typographical device is adequate for systems governed by differential equations, but in more general situations
where the  dynamical evolution  is described by mathematical flows the replacement of $t$ by $-t$ does not necessarily correspond to a physical time reversal \cite{bernstein}. The disconnect between reversions and sign changes is even more drastic in Arrow's scheme of preference rankings. In this situation it seems reasonable to associate reversions
with an interchange of preferences: instead of $A \succ B$ we presume $B \succ A$. In this sense, the reverse of the Condorcet triplet Eq.(\ref{sec1}) is
\be
A \succ C \succ B \succ A                           \label{condrev}
\ee
\begin{figure} [htb]
\centering
 \includegraphics[width=\linewidth]{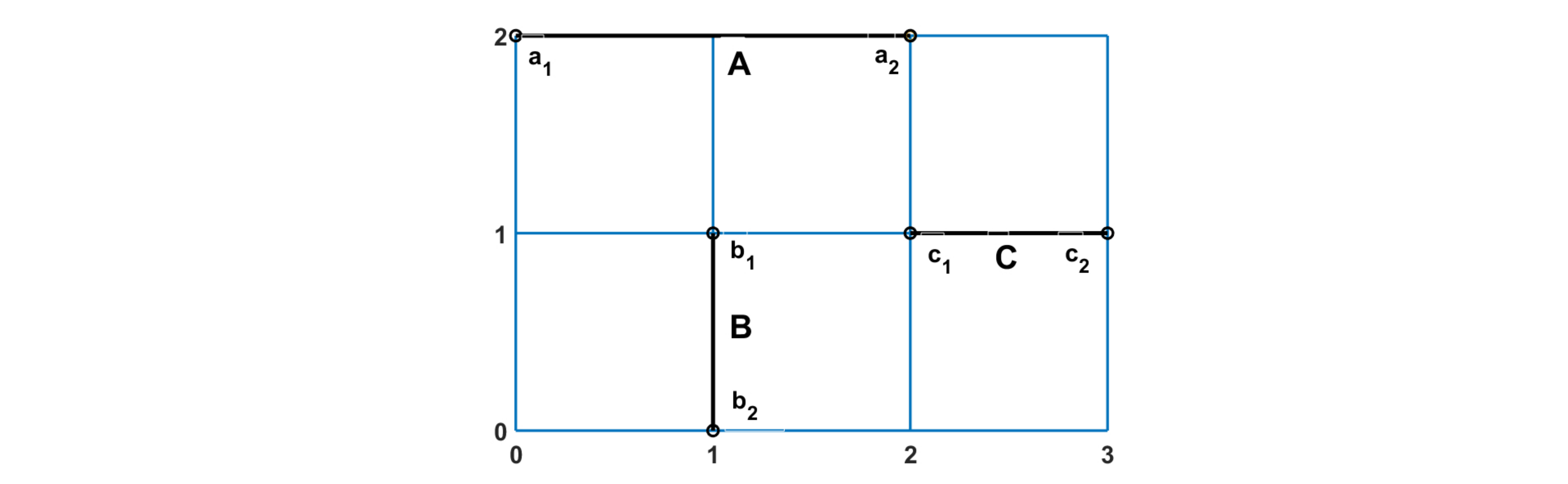}
\caption{Arrangement of three point sets $A,~B,~C$, corresponding to the Hausdorff inequalties in Eq.(\ref{unir}).}
\label{fig:3}       
\end{figure}
Switching preferences also implies a reversal of the inequalities in the corresponding Hausdorff distances : and this,
in turn, requires that the pattern of sets in Figure 2 has to be shifted. A minimal rearrangement consistent with
the sequence of preferences in Eq.(\ref{condrev}) is shown in Figure \ref{fig:3}. Comparing with Figure \ref{fig:2}, it is clear that all of
the individual points retain their positions, just the labeling of the sets has changed. The resulting collection of preferences and Hausdorff distances can then be summarized in a form similar to Eq.(\ref{uni}):
\bea
&& A \succ C ~\Leftrightarrow~ \delta_H(A \rightarrow C)=\sqrt{5}\ > \ \sqrt{2} = \delta_H(C \rightarrow A)\nonumber\\
&& C \succ B ~\Leftrightarrow~ \delta_H(C \rightarrow B)=\ 2\ \ > \ \sqrt{2} = \delta_H(B \rightarrow C)\label{unir}\\
&& B \succ A ~\Leftrightarrow~ \delta_H(B \rightarrow A)=\sqrt{5}\ > \ \sqrt{2} = \delta_H(A \rightarrow B)\nonumber
\eea
Evidently, this perspective on reversions has no relation to the introduction of minus signs.


\end{document}